\documentclass{article}
\usepackage{spconf,amsmath,graphicx}
\usepackage{amsmath}
\usepackage{amsfonts}
\usepackage{xcolor}
\usepackage{enumitem}
\setlist{nosep, leftmargin=14pt}

\usepackage{mwe} 

\usepackage{url} 

\title{Fast multi-contrast MRI using joint multiscale energy model}
\name{Nima Yaghoobi$^*$, Jyothi Rikhab Chand$^*$, Yan Chen$^*$, Steve R. Kecskemeti$^\dag$, James H. Holmes$^\ddag$, Mathews Jacob*\thanks{This work is supported by NIH grants R01-AG067078, R01-AG087159, R01-HD108868, P50 HD103556, NIH R01 HD108868 and R01-EB019961.}}
\address{$^*$University of Virginia, $^\dag$ University of Wisconsin-Madison, $^\ddag$ University of Iowa }
\begin{document}
%
\maketitle
\begin{abstract}
The acquisition of 3D multicontrast MRI data with good isotropic spatial resolution is challenged by lengthy scan times. In this work, we introduce a CNN-based multiscale energy model to learn the joint probability distribution of the multi-contrast images. The joint recovery of the contrasts from undersampled data is posed as a maximum a posteriori estimation scheme, where the learned energy serves as the prior. We use a majorize-minimize algorithm to solve the optimization scheme. The proposed model leverages the redundancies across different contrasts to improve image fidelity. The proposed scheme is observed to preserve fine details and contrast, offering sharper reconstructions compared to reconstruction methods that independently recover the contrasts. While we focus on 3D MPNRAGE acquisitions in this work, the proposed approach is generalizable to arbitrary multi-contrast settings.
\end{abstract}

\section{Introduction}

\label{sec:intro}

The superior soft tissue contrast of MRI sets it apart from other medical imaging modalities. However, the acquisition of different contrasts at high and isotropic spatial resolution often results in long scan times. 
Several methods have recently been introduced to reduce scan time in multicontrast MRI acquisitions \cite{t1shuffling,qalas}. 
For example, T1 schuffling \cite{t1shuffling} and MPnRAGE \cite{mpnrage} aims to recover multiple T1 weighted images from a single acquisition, while 3D QUALAS \cite{fujita2019three} aims to acquire T1 and T2 weighted images. Magnetic resonance fingerprinting, acquires data with numerous different weightings in a massively undersampled fashion. Many of the above schemes rely on subspace recovery, where a temporal subspace is estimated from Bloch simulations or analytical models for magnetization dynamics. The spatial factors are estimated from the measured k-t space data, often regularized using wavelet sparsity regularization. A challenge with the above subspace methods is the need for accurate models for magnetization evolution, which are often computationally demanding and may not be available in all settings. In addition, a large dictionary may be needed to capture the entire range of parameters, which translates into a large memory demand and computational complexity. The ability to compress the dictionary without losing useful information to a few basis functions is also dependent on the application. 

The main focus of this work is to introduce a joint multiscale energy model and the associated iterative reconstruction algorithm to recover multicontrast data from highly undersampled measurements. The multiscale energy (MUSE) model was originally introduced for the recovery of a single 2D image from its measurements. This approach has conceptual similarities to plug-and-play denoising models \cite{pnp} that are widely used in image recovery. These models typically require a contraction constraint on the denoiser to guarantee convergence, which often translates to lower performance. The explicit energy based formulation in MUSE enables the use of sophisticated optimization algorithms from compressed sensing literature, which offers guaranteed convergence to a local minimum, without the need for such a contraction constraint. In addition, the multiscale score-based training strategy in MUSE improves convergence to the global minimum. These desirable properties translates into improved performance, which is comparable to end-to-end trained reconstruction models \cite{modl,hammernik2018learning}. Unlike end-to-end methods that are specific to the forward model that is used during training, the learned energy model generalizes to arbitrary forward models. 

This paper extends MUSE to the joint recovery of multiple 3D volumes with different contrasts. Unlike previous multi-contrast methods \cite{t1shuffling,qalas,mrf} that rely on magnetization evolution models and joint sparsity models that use explicit edge priors, the proposed approach is data-driven. We learn the joint density, which captures the correlations between the contrasts, from the training data. In this work, we consider an inversion recovery 3D radial acquisition (MPnRAGE) \cite{mpnrage}. Here, each of the relaxation-weighted images is derived from k-space spokes acquired during a range of inversion times, making it challenging to model the contrast. We note that the magnetization changes rapidly after the inversion pulse, followed by a relatively slow change towards the end of the inversion block. We vary the number of spokes for each contrast to minimize the relaxation-induced blurring, while acquiring as much k-space information as possible. While our focus is on inversion recovery sequences, the proposed approach is general enough to be applied to other multi-contrast settings. 

Long acquisition times are needed to generate fully sampled datasets for each of the effective inversion times, increasing the risk of involuntary volunteer motion. We therefore propose to learn the model from the recovered data using a self-supervised deep factor model (DFM) algorithm \cite{dfm} using multiscale score matching. Although the DFM approach offers good recovery, the main challenge is the high computational complexity of this supervised approach (approximately 6 hours on an A100 card). The proposed supervised approach significantly reduces the computation time to approximately 12.5 minutes to recover four contrast 3D volumes with 1 mm isotropic resolution. Although unrolled supervised models can offer state-of-the-art performance, the high memory demand of these algorithms unfortunately make them infeasible in our 3D setting. This formulation enables us to use most of the optimization algorithms in the CS setting with guaranteed convergence, without requiring restrictive contraction constraints on the CNN. As shown in \cite{muse}, the multiscale approach offers a performance that closely matches unrolled models, while being agnostic to the specific forward model.


\section{Methodology}
\label{sec:method}

\subsection{Problem formulation}\vspace{-0.5em}
We consider the recovery of $C$ contrast images $\boldsymbol \gamma_i; i=1,..,c$ from their undersampled and noisy measurements:
\begin{equation}
\mathbf b_i = \mathcal A_i(\boldsymbol\gamma_i) + \mathbf n_i; ~~~ \mathbf n_i \sim \mathcal N(0, \eta^2). 
\end{equation}
Here, $\mathcal A_i$ denotes the forward model for the $i^{\rm th}$ contrast and $\mathbf n_i$ is the noise of variance $\eta^2$. The measurement process of all the images can be compactly represented as 
\begin{equation}
\mathbf B = \mathcal A(\boldsymbol\Gamma) + \mathbf N
\end{equation}
Here, the matrix $\boldsymbol\Gamma = \begin{bmatrix}\boldsymbol\gamma_1|..|\boldsymbol\gamma_{c}\end{bmatrix}$ and  $\mathcal A$ is the concatenation of the individual forward models $\mathcal A_i$, which  represents the forward model for the \(i\)-th contrast, including coil sensitivity maps and the non-uniform Fourier transform. Similarly $\mathbf B = \begin{bmatrix}\mathbf b_1|..|\mathbf b_c\end{bmatrix}$ and $\mathbf N = \begin{bmatrix}\mathbf n_1|..|\mathbf n_c\end{bmatrix}$ are the acquired k-space data of the contrasts and the corresponding noise variables. 
\subsection{MAP image recovery using joint MuSE}\vspace{-0.5em}
In this work, we model the joint probability density of the contrasts using a neural network as 
\begin{equation}
p_{\theta}(\boldsymbol \Gamma) = \frac{1}{Z_{\theta}} \exp\Big(-{\mathcal E_{\theta}(\boldsymbol \Gamma)}\Big),
\end{equation}
Here, \(\mathcal E_{\theta} \) is a CNN based joint energy model, which learns the correlation between the different  contrasts. We learn the parameters of $\mathcal E_{\theta}$ using multiscale denoising score matching \cite{muse}:
\begin{equation}
    \theta^* = \arg \min_{\theta} \left(\mathbb E_{\boldsymbol\Gamma}~\mathbb E_{\sigma}~ \mathbb E_{\mathbf Q} ~\|\nabla \mathcal E_{\theta}(\boldsymbol\Gamma + \sigma \mathbf Q)-\sigma \mathbf Q\|^2 \right),
\end{equation}
where the entries of $\mathbf Q$ are zero mean Gaussian random variables with unit variance. 

With the above formulation, the maximum aposteriori recovery of $\boldsymbol{\Gamma}$ can be posed as the minimization of the log posterior:
\begin{equation}
C(\boldsymbol \Gamma) = \frac{1}{2 \eta^2} \|{\mathcal A} \left(\boldsymbol \Gamma\right) - \mathbf B \|^2 + {\mathcal E_{\theta}(\boldsymbol \Gamma)}\Big.
\end{equation}

The learning of a 3D energy model is challenging, especially when there are few datasets to learn from. We instead propose to learn a 2D energy from coronal, sagittal, and axial slices of the 3D MRI volume:
\begin{equation}\label{sumdirections}
\mathcal E_{\theta}(\boldsymbol \Gamma) = \sum_m E_{\theta}\left[\mathcal S_{x,m}(\boldsymbol \Gamma)\right] 
 + \sum_m E_{\theta}\left[\mathcal S_{y,m}(\boldsymbol \Gamma)\right]
 + \sum_m E_{\theta}\left[\mathcal S_{z,m}(\boldsymbol \Gamma)\right]
\end{equation}
Here $\mathcal S_{x,m},\mathcal S_{y,m},\mathcal S_{z,m}$ are slice extraction operators that extracts the $m^{\rm th}$ slice along the $x,y,z$ directions, respectively and $E_{\theta}:\mathbb R^{2p\cdot c}\rightarrow \mathbb R^+$ denotes a 2D CNN-based energy model. Here, $p$ is the number of pixels in the 2D complex images and $c$ is the number of contrasts. We learn the joint energy model from coronal, sagittal, and transverse 2D slices using stochastic gradient descent, which is more efficient from a data and memory perspective.
Because the extracted slices for a specific direction do not overlap, we have $\sum_m \mathcal S_{x,m}^T \mathcal S_{x,m}( \boldsymbol \Gamma) =\boldsymbol \Gamma $, which is also true for the other directions. 

\begin{figure*}[t!]
            \centering
            \includegraphics[width=0.75\textwidth]{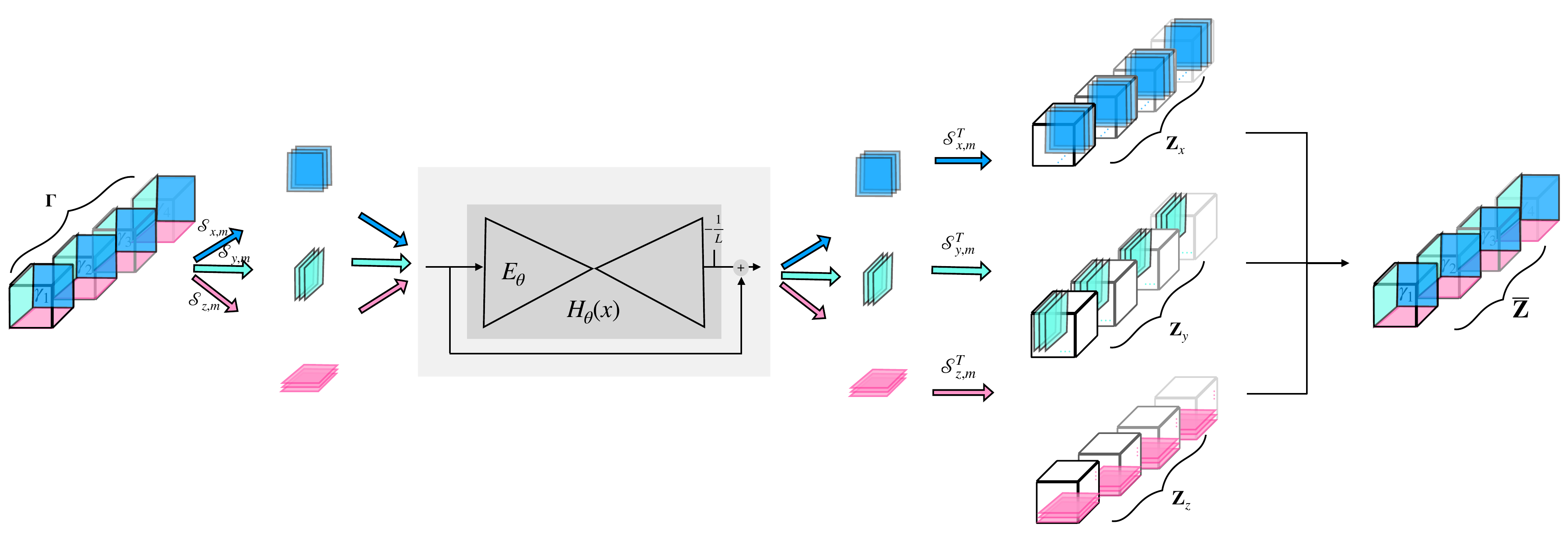}\vspace{-1em}
            \caption{Overview of the proposed approach. The goal is to recover $\boldsymbol{\Gamma}$, which is a concatenation of four contrast images. The 2D energy $E_{\theta}$ is defined in slicewise fashion, each consisting of four different contrasts that are fed as channels. The scalar energy value is indicative of the quality of the four images. We utilize $E_{\theta}$ to define $\mathcal E$ by feeding the coronal, sagittal, and transverse slices to $\mathcal E$ is a slice by slice fashion as shown in \eqref{sumdirections}. The gradient $H_{\theta}(\mathbf x) = \nabla E_{\theta}(\mathbf x)$ is used to evaluate the denoised versions $\mathbf Z_x$, $\mathbf Z_y$ and $\mathbf Z_z$ as shown in \eqref{denoised1}-\eqref{denoised3}. The image recovery algorithm proceeds by iteratively solving \eqref{cg}.
            }\vspace{-1em}
            \label{fig:mainfig0}
        \end{figure*}
        
\subsection{Majorize-Minimize Algorithm}\vspace{-0.5em}

We note that $E_{\theta}(\mathbf x)$ can be majorized \cite{muse} as 
\begin{eqnarray*}
    {E_{\theta}(\mathbf x)}&\leq&
    c+ \frac{L}{2} \left\| \mathbf x - \underbrace{\left(\mathbf x_{n}- \frac{1}{L} H_{\theta}(x_n)\right)}_{\mathbf z_n} \right\|^2,
\end{eqnarray*}
where $\mathbf x_n$ is the current iterate and $c$ is a constant. Here, $L$ is the Lipschitz constant of \( H_{\theta}(\mathbf x_{n}) \), which is the gradient of the energy model at the $n^{\rm th}$ iteration. 
Extending to \eqref{sumdirections}, we obtain
\begin{eqnarray*}
\mathcal E(\boldsymbol \Gamma) &\leq& 
 C +\frac{L}{2} \Big( \|\boldsymbol \Gamma - \mathbf Z_x\|^2 + \|\boldsymbol \Gamma - \mathbf Z_y\|^2 + \|\boldsymbol \Gamma - \mathbf Z_z\|^2\Big)
\end{eqnarray*}

Here, $C$ is a constant and 
\begin{eqnarray}
\label{denoised1}
    \mathbf Z_x &=& \sum_m \mathcal S_{x,m}^T \Big(\mathcal S_{x,m}(\boldsymbol\Gamma_n) - \frac{1}{L}H_{\theta}(\mathcal S_{x,m}(\boldsymbol\Gamma_n))\Big)\\
\label{denoised2}
    \mathbf Z_y &=& \sum_m \mathcal S_{y,m}^T \Big(\mathcal S_{y,m}(\boldsymbol\Gamma_n) - \frac{1}{L}H_{\theta}(\mathcal S_{y,m}(\boldsymbol\Gamma_n))\Big)\\
\label{denoised3}
    \mathbf Z_z &=& \sum_m \mathcal S_{z,m}^T \Big(\mathcal S_{z,m}(\boldsymbol\Gamma_n) - \frac{1}{L}H_{\theta}(\mathcal S_{z,m}(\boldsymbol\Gamma_n))\Big)
\end{eqnarray}

With the majorization, the original cost function can be majorized as 
\begin{eqnarray}
    C\left(\boldsymbol{\Gamma}\right) &\leq& \frac{\|\mathcal A(\boldsymbol\Gamma) - \mathbf B\|^2}{2\eta^2} + \frac{3L\Big( \|\boldsymbol \Gamma - \overline{\mathbf Z}\|^2 
    \Big)}{2} ,   
\end{eqnarray}
where $\overline{\mathbf Z} = \left(\mathbf Z_x+\mathbf Z_y+\mathbf Z_z\right)/3$. The above equation can be solved as 
\begin{equation}
\label{cg}
    \boldsymbol\Gamma_{n+1} = \left(\frac{\mathcal A^H\mathcal A}{\eta^2} + {3L}\mathcal I\right)^{-1}\left(\frac{\mathcal A^H\mathbf B}{\eta^2} + {3L\overline{\mathbf Z}}\right)
\end{equation}
Here, \( \mathcal A^H \) is the Hermitian transpose of the forward model \( \mathcal A \). We solve \eqref{cg} using conjugate gradient algorithm. 

We set the hyperparameter \(\eta=25\) which was determined by optimizing over the data of a subject. The Lipschitz constant $L$ is estimated as in \cite{muse} to be 1.88; we set it to 2 in our experiments to ensure monotonic convergence. With these settings, the algorithm converges in approximately 30 iterations.

\section{Implementation details}
\subsection{Sequence and Data acquisition}\vspace{-0.5em}
 MPnRAGE, a 3D radial inversion recovery sequence, is used to collect the data. With a golden angle view ordering, the sequence employs 224 inversion blocks, each of which consists of an inversion pulse, 385 radial gradient echoes spaced 4.88 ms apart, and a 503.5 ms magnetization recovery delay before the subsequent inversion pulse. 
The entire acquisition duration is 9 minutes to acquire the brain with \(1 \times 1 \times 1 \, \text{mm}^3\) resolution and a \(256 \times 256 \times 256 \, \text{mm}^3\) field of view (FOV). We considered four subsets of the readouts (TI$_1=314.56-417.04$ms, TI$_2=558.56-758.64$ms, TI$_3=758.64-1056.32$ms, and TI$_4=1583.36-1881.04$ms) to generate the contrasts. All inversion blocks were used to generate the reference data, while 56 of the 224 inversion blocks were retained to test the proposed method. This corresponds to an acceleration of 4x and a scan time of 2.25 minutes.

\begin{figure}[t!]
            \centering
            \includegraphics[width=0.5\textwidth]{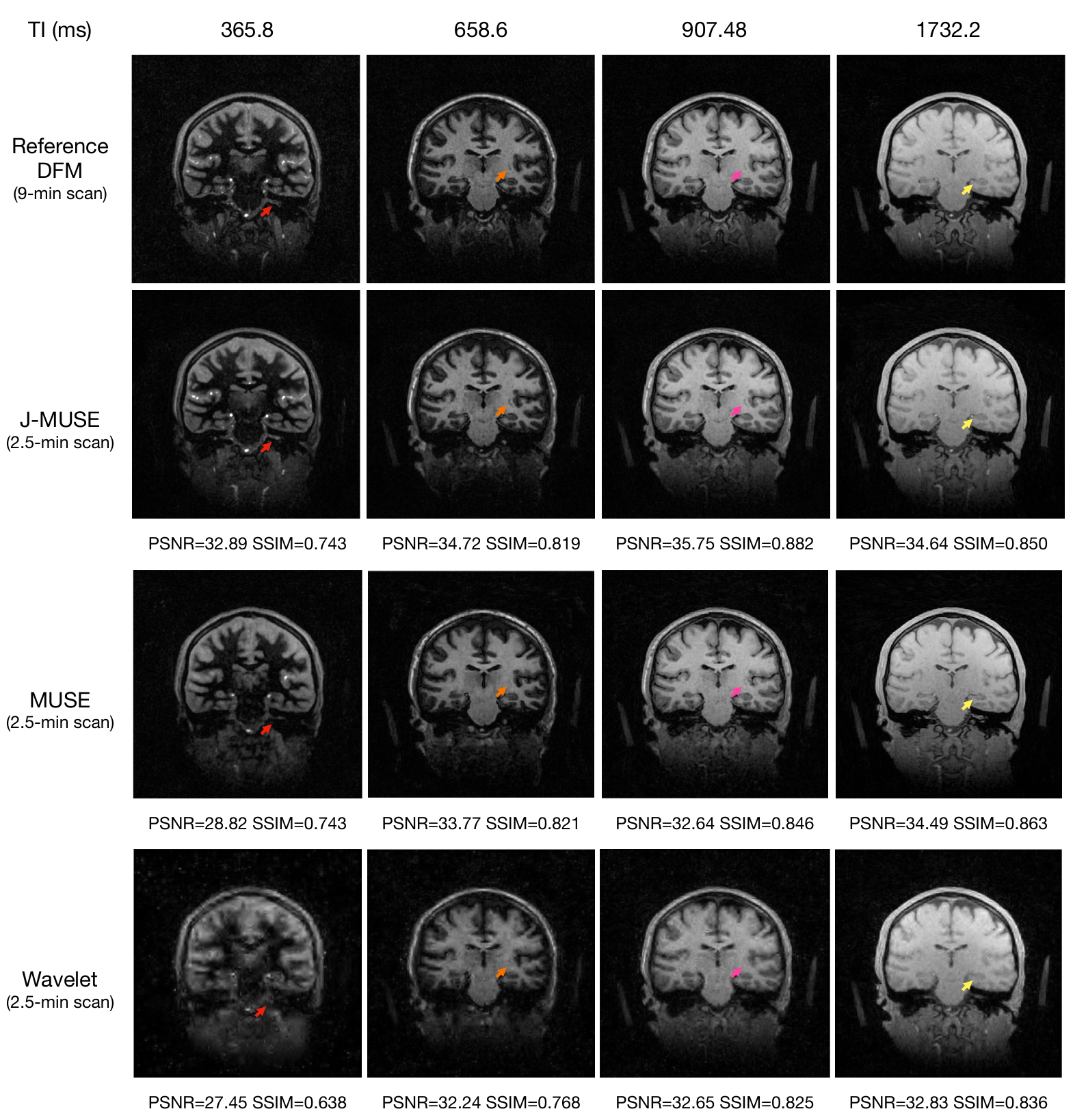}\vspace{-1em}
            \caption{Comparison of the proposed method on MPnRAGE data: The columns corresponds to different contrasts, where the average inversion time is marked as TI. 
            The first row denotes a coronal slice extracted from the reference volumes, which are recovered from the full 9 minute scan using self-supervised DFM algorithm. The reconstructions using the joint energy model (J-MUSE), conventional energy model (MUSE), which recovers each contrast independently, and wavelet regularized reconstruction (Wavelet) that also recovers each volume independently are shown. The PSNR and SSIM of the reconstructions from 2.5 minutes of data (4x undersampled compared to the reference) with the reference image is also reported.}

            \label{fig:mainfig1}
        \end{figure}

Five normal healthy volunteers (3 male and 2 female, 20-50 years of age) were imaged using a 48-channel receive RF head coil on a clinical GE-3T Premier scanner. The participants were instructed to remain still during the acquisition. Principal Component Analysis (PCA) was applied to condense the initial 48 channels of data into 4 virtual coils. The JSENSE approach was utilized to estimate the coil sensitivity maps of the virtual coils.

\subsection{Architecture and Training of Neural Network}\vspace{-0.5em}

We represent the energy function \( E_{\theta}(x) \) as 
\begin{equation}
E_{\theta}(\mathbf x) = \frac{1}{2} \|\mathbf x - \phi_{\theta}(\mathbf x)\|^2
\end{equation}
where \(\phi_{\theta}\) is parameterized through DRUNet configured with 64, 128, 256, and 512 channels, respectively. Each of these layers incorporates skip connections that link the downscaling via 2x2 strided convolutions with upscaling achieved through 2x2 transposed convolutions. 
The model processes 8 input and output channels, corresponding to 4 contrast images with real and imaginary components. 
%
%
We trained the model using denoising score matching with ADAM optimizer with a learning rate of \(1 \times 10^{-4}\) for 200 epochs on the 3D data from four subjects, while the algorithm was tested on the data from the fifth subject.

\begin{figure}[t!]
            \centering
            \includegraphics[width=0.5\textwidth]{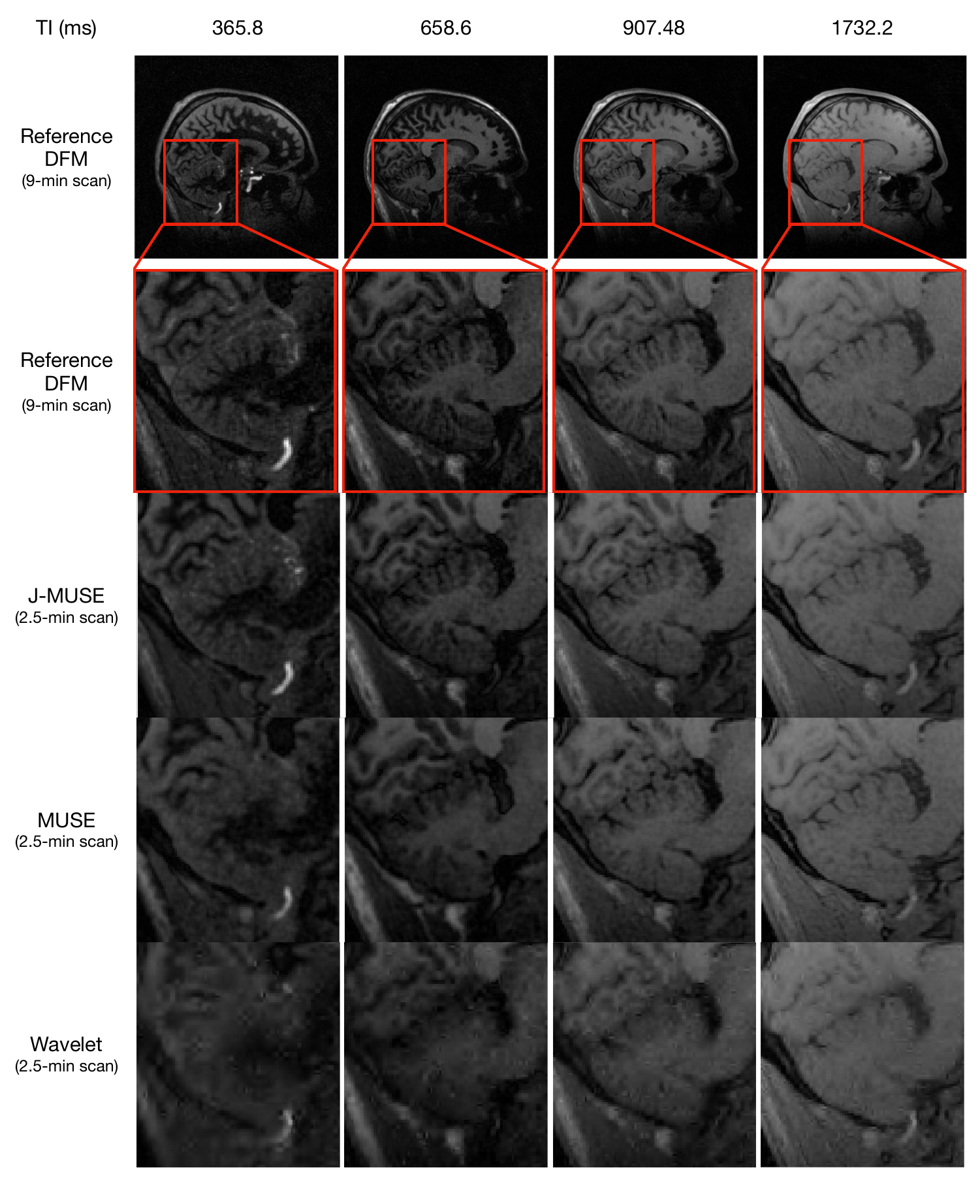}\vspace{-1em}
            \caption{Comparison of the proposed method on MPnRAGE data: The columns corresponds to a sagittal view extracted from different contrasts. The average inversion time is marked as TI. The first and second row corresponds to the slice and a zoomed section of it. Rows 3-5 correspond to J-MUSE, MUSE that recovers each volume individually, and Wavelet regularized recovery that recovers each volume independently.}\vspace{-2em}
            \label{fig:mainfig2}
        \end{figure}
        
\section{Experiments and Results}
\label{sec:res}
\vspace{-3mm}
We compare the reconstruction performance of the joint recovery approach against MUSE and wavelet-regularized recovery. Both of these algorithms reconstruct each of the contrasts independently. As shown in Fig 1 and 2, the proposed approach outperforms MUSE and wavelet recovery in terms of both PSNR and SSIM across all four TIs (365.8 ms, 658.6 ms, 907.48 ms, and 1732.2 ms). 
The proposed approach is observed to preserve fine details and structural integrity of the edges, yielding clearer and sharper reconstructions that are closer to the reference images. We note that the number of k-space spokes are significantly smaller for the first contrast (TI=365.8ms); the independent recovery approaches exhibit significant distortions in this case, both in terms of edge fidelity and contrast differences. By contrast, the joint approach is able to transfer information from other well-sampled contrasts to preserve these details. 

The improved performance and consistency of the proposed approach can be attributed to the learning of the joint probability model, which captures the similarity between different contrasts. By leveraging shared features across multiple contrasts, the proposed approach is able to achieve more accurate and reconstructions that are less noisy.


\section{Conclusion}\vspace{-3mm}
In this study, we introduced a multiscale energy-based model for the joint recovery of multiple 3D contrast volumes from undersampled measurements. The CNN model learns the joint probability density, which accounts for the complex interdependencies between the image volumes. The energy-based formulation offered reconstruction of the four 3D volumes in around 12 minutes of run time, while offering improved reconstructions than approaches that independently recover each contrast. While our focus is on 3D inversion recovery acquisition, the proposed model is applicable to arbitrary multicontrast acquisition settings.

\section{Compliance with ethical standards}\vspace{-3mm}
The data in this research study was acquired on a 3T GE Premier MRI scanner at the University of Iowa (UI). This study was approved by the institutional review board (IRB) at UI, and informed written consent was obtained from the volunteer prior to scanning.


\bibliographystyle{IEEEbib}
\bibliography{main}

\end{document}